# AI and Blackness: Towards moving beyond bias and representation

Christopher L. Dancy, *Member*, *IEEE*, and P. Khalil Saucier

*Abstract—* In this paper, we argue that AI ethics must move beyond the concepts of race-based representation and bias, and towards those that probe the deeper relations that impact how these systems are designed, developed, and deployed. Many recent discussions on ethical considerations of bias in AI systems have centered on racial bias. We contend that antiblackness in AI requires more of an examination of the ontological space that provides a foundation for the design, development, and deployment of AI systems. We examine what this contention means from the perspective of the sociocultural context in which AI systems are designed, developed, and deployed and focus on intersections with anti-Black racism (antiblackness). To bring these multiple perspectives together and show an example of antiblackness in the face of attempts at de-biasing, we discuss results from auditing an existing open-source semantic network (ConceptNet). We use this discussion to further contextualize antiblackness in design, development, and deployment of AI systems and suggest questions one may ask when attempting to combat antiblackness in AI systems.

## I. INTRODUCTION

The discussions surrounding ethical considerations of AI systems have increased in recent years, with both academic and popular publication venues used to discuss a wider range of ethical issues surrounding the design, development, and deployment of AI systems; this includes the creation of the ACM conference on Fairness, Accountability, and Transparency (ACM FAccT, in 2018), as well as the AAAI/ACM conference on Artificial Intelligence, Ethics, and Society (AIES in 2018). Some of these ethical considerations have revolved around the topic of race and corresponding bias (e.g., see [1-3]).

However, a focus on racial bias has too often taken a more individualistic perspective and overlooked a deeper ontological issue with the artifact of racialization and its relation to the dominant representation of *the Human* (i.e., the Human as discussed by Wynter [4]). This deep, structural issue is key to understanding how we may create ethical AI systems that are fairer and more equitable as it forces us to contend with the episteme that drives the actions of the designer, developer, and deployer of an AI system (who may be one person or multiple people), and that helps define the learning and action space of the AI system itself.

Simon [5] defines the engineer's job, and more generally the designer, as one of synthesis, focused on designing and constructing artifacts that attain goals and serve a function. Ideally, they focus on how things "ought to be" [5]. Simon holds three aspects important in an artifact's adaptation to such a goal: the goal itself, the inner environment of the artifact, and the outer environment in which that artifact must operate.

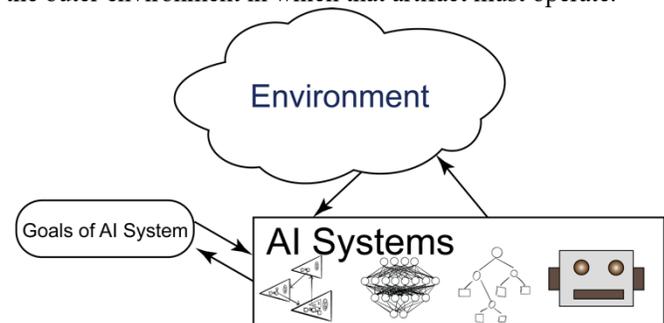

**Fig. 1.** Components important for an artifacts adaptation to a goal that are described by Simon [5].

When we consider how to create *intelligent* artifacts (i.e., intelligent systems or AI systems), we often find ourselves preoccupied and constrained by the inner environment of those systems. To get to the goal (how things "ought to be") we create some system based on our own knowledge and beliefs, which help us not only design the artifact itself, but also decide the specifics of the goal and a broad representation of the environment in which we expect the intelligent system to operate. Many AI system designers have moved to using large sources of data to define the environment in which AI systems operate to pursue those designer-defined goals. However, with the use of those data comes the assumptions of those data (O'Neil [6] discusses some of these assumptions, while Caliskan, Bryson and Narayanan [7], discuss some human-like biases that are created when developing semantic-based systems from a corpora). This puts designers, developers, and deployers at a precarious position as we must contend with potential issues at the intersection of AI, racialization, and ethics (particularly we focus on that which is anchored by antiblackness) from all three fronts discussed by Simon [5]; Fig.

---

Christopher L. Dancy is with the Computer Science Department and affiliated with the Critical Black Studies Department at Bucknell University, Lewisburg, PA 17837 USA (email: christopher.dancy@bucknell.edu).

P. Khalil Saucier is Chair of the Critical Black Studies Department at Bucknell University, Lewisburg, PA 17837 USA (email: khalil.saucier@bucknell.edu).

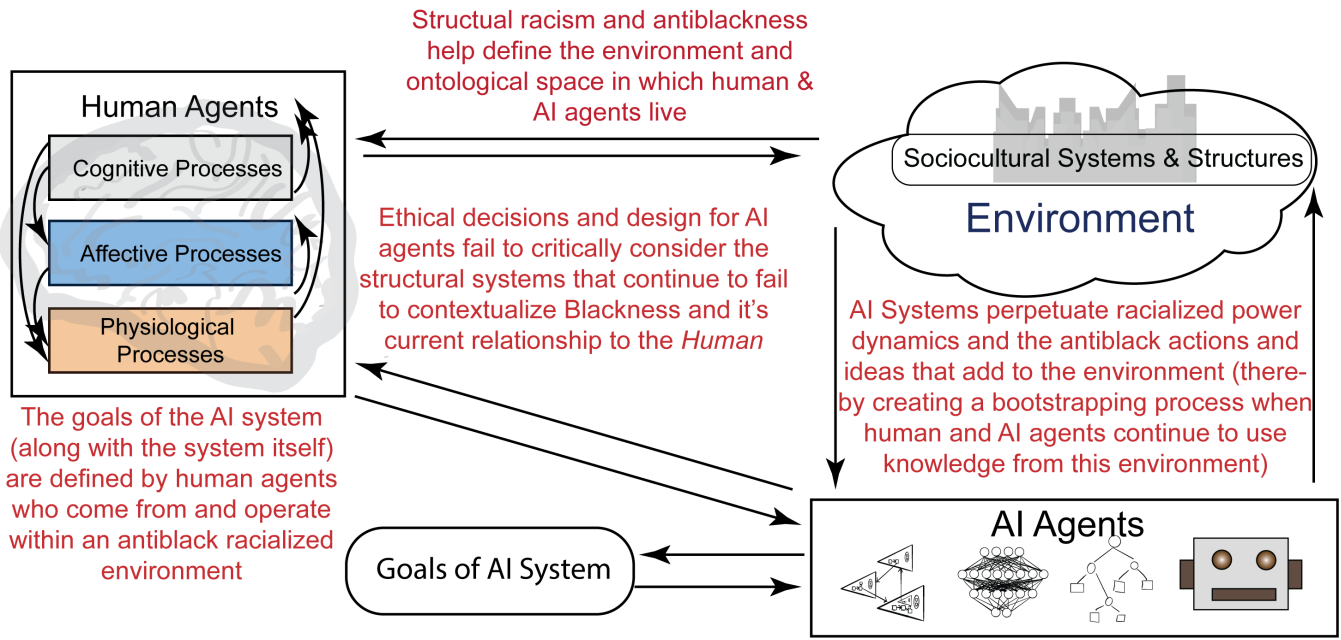

**Fig. 2.** Components important for an intelligent artifact's adaptation to a goal in the context of Blackness and the ethics of developing those artifacts. We contextualize these points individually within related sections below.

2 shows some of these intersections. That is, we must contend with the internal knowledge system of the designer or engineer and the outer environment that provides contextual training and testing of those systems. In other words, it is essential that we reckon with the contextual and paradigmatic that is incarnate within AI systems. We must grapple with the political ontology of anti-Black racism and race that defines much of our actions, modeling, and more.

We maintain that there is a need to rethink the current paradigm for understanding ethics in AI as they relate to race, and that a good place to begin is with (pervasive) antiblackness and AI systems. To do this, we must start by exploring *the Human*, for race is an entirely social artifact built-upon and driven by historical (and current) power dynamics, as well as self-interest. Thus, there is a need to work at the intersection of AI and Black Studies in a way that contextualizes *Blackness* and *the Human* in the face of the design, development, and use of AI systems. This paper focuses particularly on Blackness, because of the pervasiveness of antiblackness across sociocultural systems and its contrast with whiteness, which is used (by default) to define the Human, and thus often the *goal outputs* of AI systems. We must attend to the contexts from which AI is enunciated and created, which means to attend to the very conditions of knowledge itself which is at the core of western modernity. The history of modernity is the history of the west and its monopolization of the Human.

In the following sections, we discuss relations between Blackness, antiblackness, and the Human as a way to engage with the historical (and current) sociocultural structures that enable (and in some ways perform) antiblackness. We then discuss the different portions of Fig. 2 in separate sections, separating into the structural systems undergirded by antiblackness as well as Blackness in AI ethics. To make things more concrete, when we use terms like racism or racist, we tend to focus on anti-Black racism and racist, though the understanding of Blackness is important to understanding other forms of racism.

## II. BLACKNESS AND THE HUMAN

To ethically engage in a conversation between AI and race (and thus Blackness) requires moving beyond racial representation, bias, and discrimination. That is, it requires an understanding of antiblackness as a structuring principle for thought and action, which is enacted using concepts of race. In other words, antiblackness structures not just outcomes (e.g., racial inequality, medical apartheid, and more), but structures practices of knowing, thinking, and modeling. Just as inequality is a symptom of antiblackness, so too, it can be argued, is much of AI design (e.g., many of the systems that have been developed and deployed have enabled particularly inequitable and anti-Black practices, [2]).

The distinction between human intelligence and artificial intelligence is enabled by a belief in an onto-epistemological structure that positions Blackness as the negation of the Human or at the very least the infrahuman, as the index against which the Human is measured and constructed/designed. Blackness, in this sense, is the name given to the antagonism of the Human world, that which helps its material configuration and articulation. This configuration and articulation is key to how we define human intelligence and thus how we define artificial intelligence that is synthesized [5] by the Human. Thus, to explore the entanglement of Blackness within AI, is to understand Blackness beyond identity and category and understand it as:

(1) a sign, symbol, or metaphor that represents something,

namely lack and dereliction [8]; and

(2) as a paradigmatic expression of the world.

Put differently, the Human is predicated on a regime of social organization that excludes Blackness. AI is configured along the political ontology of race in that it relies on the entwined nature of the Human and non-human (or *Other*). Blackness reinscribes the Human as synonymous with *western Man* [4]. The Human marks the agentic properties for all AI design, not Blackness. *Human* intelligence is reliant on the dominant, standard *genre of the Human [9]* that positions the "Western bourgeoise liberal monohumanist" as the Human. This requires the "wholly-Other" status of those of African descent, that is Blackness, as well as the descriptive codes or memes that specify the "symbolic life/death" of homo oeconomicus [9, pp 29]. To draw the line between *human* intelligence and *artificial* intelligence, there must be an account for *naturally* human and artificial. That *natural* definition of the Human is reliant on our present sociogenic "Darwinian *descriptive statement*" [9, pp 29] that informs and undergirds what it means to be a *natural* human and by consequence what it means to be artificial, or, according to Simon [5, pp 5], "synthesized…by *human beings*" and "imitate appearances in *natural* things while lacking the reality of the latter" (emphasis ours).

Indeed, even the use of intelligence as a conceptual marker brings with it a certain "value-laden" history [10]. The typical account for intelligence is reliant on historically eugenic science that used racial hierarchy as a structuring principle. Thus, to think and act ethically about Blackness (and by extension, race) and AI, we must recognize the degree to which the political ontology of Blackness affects and structures the world; that is, we must take into consideration the innocuousness ascribed to the Human.

### III. STRUCTURAL SYSTEMS, ANTIBLACKNESS, AND AI

While many continue to focus on individuals and specific acts or bias towards anti-Black, racist outcomes (e.g., concepts related to *implicit bias*, [11-13], and individual acts of *dehumanization*, [14]), this individual focus obscures a structural cause that historically has had a longer lasting and more pervasive impact [15, 16]. Structural racism, enacted through policies at social institutions above the level of the individual, acts as the environment in which AI systems are designed, developed, and deployed. Structural racism is also enacted through the sociocultural knowledge and systems created as a result of institutional policies (some of which might be considered more formal *dogma*, while other knowledge may take the less formal or more relaxed form of *doxa* [17]).

These structural forms of racism result in anti-Black outcomes: they result in disparities in healthcare treatment [18], the assignment of criminality [19, 20], the continued segregation of resources [21, 22], and a plethora of other anti-Black outcomes (e.g., [23-25]). These outcomes further result in modification of knowledge systems that also inform structural policies, creating a pernicious cycle of implicit and explicit antiblackness.

Noble [24] audits and observes the ways in which the Google search engine treats Blackness and women, particularly focusing on the intersection thereof. She notes that "search engine results perpetuate particular narratives that reflect historically uneven distributions of power in society" ([24], pp. 71). Noble uses Google because it is a "broker of cultural imperialism" ([24], pp. 86). She uses the "pornification" of Black women and girls within the Google search engine to shed light on the commodified nature of the seemingly neutral results returned. Search engine results here are not neutral, but instead a reproduction of sociocultural knowledge for the purposes of advertising as well as the collection of more "behavioral surplus" data [26], like personalized click-through data (e.g., [27]), all factors used to learn which results to return over-time.

The example of audited Google search results is a particularly useful example when thinking about the modification of knowledge systems as one considers the growth of Google (and similar search engines) as an ever-growing important arbiter of knowledge. Sparrow, Liu and Wegner [28] discuss the effects of having information available via systems like internet search engines on transactive memory (which can be thought of as a "combination of memory stores held directly by individuals and the memory stores they can access because they know someone who knows that information" [28]). They found that knowing that such information is available via a computational system like a search engine reduced the likelihood of (an individual) storing specific information about that item to memory and instead resulted in storing information on where to find that item in memory. Sparrow, Liu and Wegner [28] contend that we have become dependent on these systems in the same way that "we are dependent on all the knowledge we gain from our friends and coworkers—and lose if they are out of touch." Thus, that AI systems sort and decide knowledge to give us is important given that these systems pull information from structurally anti-Black environments and because we treat the information retrieved in a manner similar to the ways we've treated knowledge from some friends and family in the past — they create a bootstrapping process that reflects existing anti-Black structures and enacts cycles of those structures through the decisions made.

The pervasive use of large amounts of data to inform AI systems has resulted in a more recent push towards examining the ways one might change those AI systems or those data for less biased (which often includes some form of racial bias) outcomes in the use of those systems or data or both [1, 7, 29-31]. Despite general forms of bias reduction in systems, dehumanization of Blackness persists. To show an example of this persistence despite measures of bias reduction, we examine the ConceptNet semantic knowledge network.

#### A. ConceptNet

ConceptNet [32] is an open-source knowledge graph that includes knowledge from several sources to connect *terms* with "labeled, weighted edges (assertions)." Based on the several knowledge sources, the ConceptNet API uses a system called

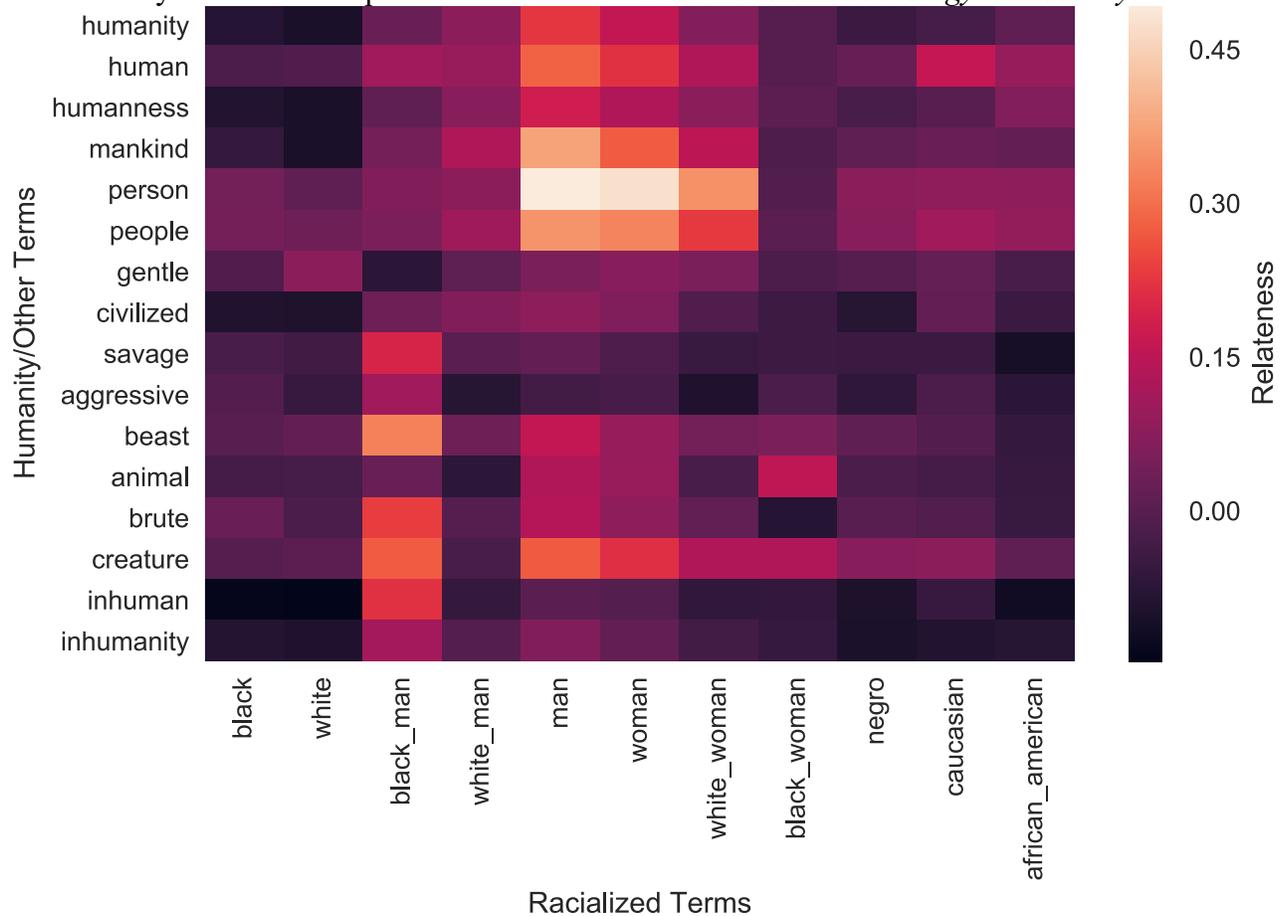

**Fig. 3.** A relatedness heatmap of human/sub-human and black/white related concepts. The relatedness values were obtained from ConceptNet 5.7. Appendix B contains a table with the values used to populate this heatmap.

"ConceptNet Numberbatch" to compute a *relatedness* between terms. ConceptNet Numberbatch is a set of word embeddings that can be used for tasks such as representing the meanings of words. The system combines data from several sources, including ConceptNet 5, word2vec, GloVe, and OpenSubtitles 2016. ConceptNet Numberbatch uses a particular algorithm to calculate the relatedness for out-of-vocabulary terms (this is useful to know given that one of the terms mentioned in an ensuing section uses this algorithm):

(1) If the word is not English, attempt to find an equivalently-spelled word within the English portion of the embeddings

(2) If the word is in English, remove a letter from the end, and see if that is a prefix of known words. If this is the case, compute the average the embeddings of those known words

(3) If the prefix is still unknown after removing one letter, continue to remove letters from the end until either a known prefix is found or you reach a single character

A more recent version of this system has also attempted to reduce *bias* (i.e., in the sense of *gender* or *racial* bias) in the system and thus produce more *fair* relations between terms through "algorithmic de-biasing" (see [31, 33], for a background on approaches used to reduce bias). The various knowledge sources used and the already completed "de-biasing" makes ConceptNet an interesting site for exploration into the pervasiveness of antiblackness within existing online knowledge sources, which can be thought of as providing a picture into the social knowledge structures that are used by humans and AI systems alike; including considering what those knowledge systems look like after bias has been considered.

As an initial exploration into how ConceptNet might be used to define *relations* between Blackness (and some related concepts) and Humanity, we used the API to explore relatedness between certain concepts; this exploration was performed using ConceptNet 5.7. We chose not to use the typical derogatory (US-centric) terms used to represent black (or indeed white) people as these would be less material for this particular focus, because the power dynamics, structural inequalities, and racist representations that are used to justify those hierarchies and inequalities are embedded within the terms used. Not only are these the terms used more commonly in everyday language, these are the terms more likely to be used and referenced for decision-making and action selection by AI systems. We used existing historical accounts of antiblackness and related racialization [4, 16, 19] as a guide for terms to use in this audit.

Fig. 3 shows a heatmap of the relatedness between race terms

(particularly some used for black or white racial categories) and terms associated with *human* (and some associated with terms meant to represent sub or other-than human); a table of these values can be found in Appendix B. As one may expect, the results are mixed when comparing terms that were particularly unbalanced in the number of edges connected to those terms. The terms associated with the _*man* and _*woman* columns illustrate this point. The term *black_man* has a documentation of 6 edges while *white_man* has 41 edges, and the term *black_woman* has 0 edges (meaning that the system had to calculate those relatedness values using a method for determining relatedness for *out-of-vocabulary* terms), while *white_woman* has 9 edges; by contrast, *black, white, man, and woman* had edge totals greater than 1000.

Despite *black_man* and *white_woman* being close to one another in number of edges, *white_woman* has a much closer match in relatedness to the more general *woman*. Whereas both *man* and *woman* (and *white_woman*) show a relatedness to several of the terms, black_man shows a higher relatedness to the less-than-human concepts. The term *black_woman* generally shows a lower relatedness to all of the Human and less-than-human terms.

Conceptnet provides but one example of a potential issue with AI systems that rely on increasingly large datasets to form knowledge to be used in decision-making. As discussed by [46], these data pulled from several sources often represent the "voices of people most likely to hew to a hegemonic viewpoint". By using datasets that automatically pull primarily from these internet sources, AI systems learn from (and contribute to) data that implicitly furthers the dominant mode of the Human; this occurs both through the filtering of internet sources and through the use of the internet itself as the main source [46].

Though antiblackness certainly is not limited to English speakers (indeed see, for example, discussions by [53] and [56] for Italian, French, and other non-English contexts) the primary use of US English and UK English also means that the "Western bourgeois liberal monohumanist" [9] is overrepresented as the Human, and thus "white supremacist, misogynistic,..., etc. views" [46] related to that formulation are dominant within those data. This tie to language is important – as Fanon [8, pp 2] notes, "A [person] who possesses a language possesses as an indirect consequence the world expressed and implied by this language". AI systems that engage in these learning processes (and by action also may contribute to those same internet mediated data sources) further extend ways in which the preclusion of Blackness is used to bear "the weight of civilization." [8, pp 2].

Beyond just the raw values, these knowledge networks also provide an opportunity to contextualize not only AI systems that may potentially use this semantic network, but also a way to contextualize human behavior. One might use the distinctions found in the network to think through human decision-making during potential design, development, and deployment stages. Particularly, it is useful to probe more into what the Human, other-than-human distinctions found in the results means for those decisions.

Auditing ConceptNet and similar systems provides an opportunity to probe (digital) relations between Blackness and the Human. ConceptNet as a common-sense knowledge system can give both other AI systems and people a computational model (albeit an imperfect one, as models are) of ontology, that is, of relations in the (digital) world. Understanding these relations as they exist and as they evolve is important for understanding Blackness and AI.

## IV. BLACKNESS IN AI ETHICS

Any attempt to ethically explore the entanglement between AI and antiblackness requires that we attend to the paradigmatic and structural, while acknowledging the agentic power of individuals in developing artifacts that perpetuate and fortify this structure. Acknowledging the anti-Black structure of the Human as *the* ethical problem is important to any and all discussions about bias, race, and antiblackness in AI systems. This is to say that ethics and the applications thereof for AI systems presupposes "relations of recognition" of humanity and subjectivity – something that the "spectre of slavery destroys" [42, pp 54]; and as we discuss in a following paragraph, current codes of ethics, ethical guidelines, etc. are lacking in their attempts to explicitly address the issue of antiblackness in AI systems. To move beyond projects in AI ethics that force a certain assimilation and colonization (Benjamin [23, pp 176] provides a related discussion on design as a *colonizing project*), antiblackness and the foundations of this continuously maintained boundary [25] must be directly addressed.

Despite this importance, AI ethics principles and frameworks continue to leave out antiblackness as paradigmatic, regardless of increasing awareness of bias and racism writ large. The previously mentioned de-biasing for the Conceptnet Numberbatch is a useful example of the issues one can encounter when not more directly considering antiblackness. Our audit of that system for *semantic relatedness* between the racialized terms of *black_man*, *white_man*, *black_woman*, *white_man* (as well as standardized representations of *man* and *woman*) showed a pattern reflective of known historical relations between *human*-related representations, *animal*-related representations, and existing racial structures (portraying a social order or hierarchy). Those data showed that despite the aforementioned de-biasing of the ConceptNet Numberbatch system, *black_man* retained a closer semantic relatedness to animalistic terms, while *white_man* retained a closer semantic relatedness to terms that portray humanity. What's more, the lack of semantic representation for *black_woman* (while *white_woman* remained a stand-in for *woman*) shows the continued issue of intersectionality [34], that is the issue of racialized, gendered intersections. The *erasure* of *black_woman* in a system that has been "de-biased" is notable.

In the same way that Saucier [42, pp 56] argues that those in ethnography believe (incorrectly) that "they can attend to social life in the midst of urban decay, the militarization of schools, police brutality,...and the like without coming to terms with the political ontology of blackness", so to do AI ethics discussions, principles, and frameworks skip over addressing Blackness and antiblackness directly. Indeed, this issue of not tackling antiblackness head-on when considering AI bias and ethics can be seen in the general literature. For example, Jobin, Ienca and Vayena [35] give a picture of the "global landscape" of AI ethics guidelines and principles. While they reference 64 references in the "justice, fairness, and equity" section of that review, none of those references reference antiblackness (0) and only 14% of those references (9) even provide *some* discussion on racism or racist policy (and this is with a fairly low bar of providing some discussion or example of race/racism beyond listing the term *race* alongside other categories that are the result of forms of structural oppression, like gender or ethnicity.) General frameworks that call for development of systems to mitigate *bias* (e.g., [36]), or general mapping of ethics debates (e.g., [37]) will continue to fail to adequately address systems of oppression without more directly and explicitly focusing on these systems.

*A.   AI and the boundary of Blackness*

We cannot address ethical considerations of race within AI systems until we explicitly address antiblackness as the paradigm that shapes the lived experience of Blackness. For example, the lived experience of Blackness has long been one of routinized surveillance. In tracing the surveillance of Blackness from the transatlantic slave trade to today, Brown [25, pp 7] notes that Sociogeny [8] can be understood as "the organizational framework of our present human condition...that fixes and frames blackness as an object of surveillance." They use British military ledger *the Book of Negros*, and lantern laws of colonial New York City to focus on the surveillance of racialized Black bodies (using the technologies of the day), as well as to question the use of archive to reproduce violence (see [47] for a discussion related to this latter question). Browne argues that this reproduction of violence can be a form of boundary maintenance for Blackness, what they call "Black Luminosity" [25, pp 67]. They contend that these technologies of seeing "sought to render the subject outside of the category of the Human, un-visible" [25, pp 68]. This conception of seeing not to recognize as a part of the Human, but instead to render Blackness and thus some people as un-visible is important for considering how and why we "de-bias" AI systems.

This point is all the more apparent as we consider how AI systems have been used to widen automated surveillance and policing more recently [38]. Despite the continued scrutiny of Black people, AI systems centered on facial detection and classification have struggled to correctly identify and classify Black faces. This inability to identify can be connected to the previously mentioned concept of being un-visible. Buolamwini and Gebru [1] studied these "failures" in several commercial facial recognition systems and showed the need for more inclusive datasets to increase fairness.

However, increasing fairness in these systems may only exacerbate the routinized surveillance to which Black people are subjected (see also, [54] for a discussion on how the connections between faces and certain categories, like criminal, can be thought of an *illusion* in of itself). This forces us to question how the concept of fairness in AI may be used to intensify antiblackness. The increased opportunity for recognition by facial recognition systems does not solve the issue of un-visibility, but instead morphs the ways in which un-visibility may be manifested in the AI systems that use facial recognition. The recognizing of a face does not preclude the "boundary maintenance" [25] needed to define our current genre of the Human [9].

This surveillance example points to an issue with many of the current approaches to intersections between race, ethics, and AI systems: they attack the symptoms (or, perhaps, the features). It is not necessarily that Black faces may not be correctly classified or recognized by some AI systems, that AI systems may have a disparate impact on Black people in their quest for a home loan, or really whether we're correctly judging fairness in AI systems [39], it is that these systems are being implemented within and replacing existing machinery used in societal structures that perpetuate violence and power dynamics that enforce epidermalization [8] and thus feed antiblackness. When it is not in the self-interest of non-Black people as a group to make systems fairer for Black people, we've seen historically that self-interest will drive anti-Black decisions (e.g., see the many examples in [16]). This fact is why there have been several calls for a more representative population of developers and designers [2]. Nonetheless, an increase of Black individuals within existing structures that are built-upon, and ultimately have a self-interest goal to perpetuate, antiblackness will not guarantee the development and design of non-anti-Black AI systems (e.g., see [40], for a historical example of how an increase in individuals as police did not and has not cured the anti-Black actions of police).

What's more, we must attend to the ways in which humanity not only evades Blackness, but also the ways in which humanity and agency may be used to continue antiblackness. For example, Hartman [48] discusses the transition from slavery to a post-slavery period in which the narrative of the abolition of slavery, and the supposed agency and freedom now given to Black individuals, served to reify many of the same structural conditions seen before such an abolition. They discuss the entanglement of slavery and freedom, noting that "emergent forms of involuntary servitude, the coercive control of black labor, and the social intercourse of everyday life" during the post-slavery reconstruction era "revealed the entanglements of slavery and freedom" [48, pp 151].

Carrying these entanglements forward to thinking through AI ethics, we must explicitly think through how those *scenes of subjection* persist in the design and interaction with AI systems. We might, for example, consider how the supposed agency and freedom in healthcare decisions by Black people is limited by current AI systems used in the automation of assessment [49] that exist within current healthcare systems and that rely on boundaries of the Human, and consider relations to the lack of agency by Black slaves when it came to their health [50]. We might also consider how the agency or freedom given through the implementations of these systems often serve to reinscribe coercion of Black labor through an automation of those processes that serve to police and maintain boundaries of Blackness (for example some of those automations noted by Eubanks [38].) That is, in many instances these AI systems have served to automate and maintain the modes of knowing, quantifications, and processes that bear the spectre of slavery.

To act ethically in AI design, development, and deployment is to acknowledge the structural position from which the Human, and by extension AI, emerges. AI designers, developers, and deployers must move beyond the frame of implicit bias, and actively work to identify and combat antiblackness; without this, forms of *inclusion* may result in "unwanted exposure" within existing anti-Black systems ([23], and see also [25], as well as [22], for a related concept of *predatory inclusion*). Furthermore, our ethics must ensure that we not only account for the adaptation of our AI systems, but the adaptations that continue to occur that enable certain social systems, which are built on the racial hierarchy and racial supremacy of antiblackness, to continue to thrive.

We must also consider how narratives and logics tied to progress in AI system design and use are related to the previously discussed notion of the Human, the rational homo oeconomicus [9]. For example, Brock [51] begins to disentangle the Human from technoculture (particularly by specifying what he calls *Black Technoculture*.) Though much work in AI ethics appear to attend to increasing capacity to model aspects of "reality" without bias, he asserts that "neither algorithms nor big data sufficiently model or account for the cultural qualities that are inherent to their design, leading to ethical and moral problems" [51, pp 224]. Brock discusses the need for a technocultural matrix (referencing Dinerstein's formulation [52]) to explicitly incorporate antiblackness. They suggest Blackness to be a key category to an updated matrix, to allow the recognition of "what makes Black folk different." [51, pp 229] in a manner that also pushes back against deficit narratives or aims of respectability related to Black technology use.

*B. Questions to consider*

Thus, if researchers are to attempt to address antiblackness in their AI systems in a way that moves beyond racial *bias* and *identity* in their design, development, and deployment of AI systems, there are several foundational questions at which they might begin for their decision-making processes:

- *Who will the intended system interact with (users and those affected by the use of), on what time scales[1], and how have these users traditionally interacted with anti-Black policies or behaviors?*

- *Why are we building this system, why are these data being used organized in a certain way? How might these goals and intentions interact with existing anti-Black systems?*

- *Where in time (including historical) and space might the AI system connect to anti-Black issues representative of structural or institutional issues?*

- *What are the building blocks of the AI system and how might these blocks be implicitly anti-Black?*

- *Does my AI system require antiblackness to be "equitable"? Does it require it to function?*

This list could go on, but using this current list would, in our estimation, be a start to more critical assessment of the AI systems designed, developed, and deployed so that they may cause less enaction and performance of a particular representation of the Human that is contextualized on racialized others. This critical assessment must be seen as an ongoing *process* that must continue iteratively as one designs, develops, and deploys any AI system. We must explicitly and actively engage the current dominant social organization (that conceives of the Human in a way that excludes Blackness) to ethically address how we create and use AI systems within sociocultural systems as any ethical code will presume a recognition of humanity [42].

In addition to assessing how one's AI system may intersect with Blackness through its enaction of antiblackness, it's also useful to potentially consider how such systems may be used to unearth forms of antiblackness (though while still acknowledging the potential costs of reproducing violence [47]). It is worth considering how we may "retool" and "reimagine" these AI systems in ways that push against dominant representations of the Human and that takes "narrative seriously as a liberating tool" [23, pp 195]. Nonetheless, we must do this with the idea in mind that these ends can be used as a tool "to keep the oppressed occupied with the master's concerns" [57].

Using these systems to assess existing antiblackness also face another challenge — changes to those systems due to concerns of "bias" can effectively whitewash potentially historically useful commonsense knowledge without critically examining the effects of those systems before the change. While these changes may be helpful in future iterations of an AI system and any derivations thereof, without very explicitly noting the specific changes, and critically understanding how this change might have downstream effects on other systems, we lose out on critical opportunities to understand how these systems have changed overtime and a potentially very useful historical record of digital knowledge. That is to say, developers of these systems must begin to think much more thoughtfully about ways to document these changes so that this historical record remains intact and the effects of those changes are well documented. Without this critical assessment, we risk not truly understanding how AI systems have changed over time, particularly how and why AI systems may interact with structures of antiblackness in different ways overtime.

---

[1] *For example, the concept of time-scale based bands of behavior described in [41].*

## V. CONCLUSION

It may be tempting to assume that we have started to move past normative concepts of race and racism and need not worry about much beyond the small implicit biases that modulate behavior. This idea ignores the antiblackness upon which many scientific, technical, and social systems are built. These systems continue to adapt within a steady state of antiblackness. While there have been recent discussions of a resurgence of race-based science [43], this is just pointing out an inevitable outcome given our past approaches to race more generally [44] and Blackness more specifically.

If we are going to address race-based ethical issues in the development and design of AI systems, we must move beyond using largely color-blind ideologies and expand beyond concepts such as implicit bias (e.g., see [45] for a discussion of issues with the concept of implicit bias from the perspective of law). We must recognize that AI conceals an ontological project of devastating destruction, the extreme divide between the Human and Blackness. Only then can we ethically confront the antiblackness that haunts AI systems and create spheres of intelligence beyond *the Human*.

## APPENDIX A

Table 1. Ethical guides & principles referenced by Jobin et al. [35]. Percentages add up to 99% due to rounding errors.

| Category | Reference #s (from Jobin et al., 2019) | Percentage of *justice, fairness, and equity section* in Jobin et al. (2019) |
|---|---|---|
| No mention of bias | 48, 64, 74, 83, 100, 108 | 9% |
| Bias mentioned but without notable reference to Race | 40, 42, 45, 55, 57-59, 62, 63, 67-68, 71, 73, 77, 81-82, 86-87, 89-92, 94-98, 101, 103, 105, 11 | 48% |
| Bias and Race mentioned but without substantive discussion | 44, 46, 50, 53, 54, 60, 65, 69, 75-76, 79-80, 84, 93, 102, 104, 109, 112 | 28% |
| Bias and Race mentioned with some discussion, but without discussion of antiblackness | 52, 56, 61, 72, 85, 99, 106, 107, 110 | 14% |
| Antiblackness discussed | (None) | 0 |

## APPENDIX B: TABLES OF RELATEDNESS VALUES

Table 2. Relatedness (from ConceptNet 5.7) between Black/White concepts and human (top)/Non-human (bottom) concepts

| | HUMANITY | HUMAN | HUMANNESS | MANKIND | PERSON | PEOPLE | **GENTLE** | **CIVILIZED** |
|---|---|---|---|---|---|---|---|---|
| BLACK | -0.083 | -0.019 | -0.088 | -0.059 | 0.042 | 0.044 | -0.012 | -0.09 |
| WHITE | -0.102 | -0.013 | -0.104 | -0.102 | 0.012 | 0.032 | 0.079 | -0.096 |
| BLACK_MAN | 0.027 | 0.11 | 0.011 | 0.045 | 0.063 | 0.055 | -0.072 | 0.032 |
| WHITE_MAN | 0.077 | 0.097 | 0.072 | 0.131 | 0.08 | 0.104 | 0.01 | 0.062 |
| MAN | 0.225 | 0.282 | 0.179 | 0.376 | 0.494 | 0.352 | 0.051 | 0.083 |
| WOMAN | 0.16 | 0.218 | 0.129 | 0.275 | 0.474 | 0.329 | 0.071 | 0.059 |
| WHITE_WOMAN | 0.068 | 0.132 | 0.075 | 0.151 | 0.347 | 0.229 | 0.052 | -0.012 |
| BLACK_WOMAN | -0.002 | -0.003 | 0.007 | -0.016 | -0.01 | 0.004 | -0.017 | -0.043 |
| NEGRO | -0.048 | 0.021 | -0.022 | 0.011 | 0.074 | 0.072 | -0.002 | -0.08 |
| CAUCASIAN | -0.029 | 0.161 | 0.003 | 0.026 | 0.085 | 0.109 | 0.019 | 0.018 |
| AFRICAN_AMERICAN | 0.012 | 0.095 | 0.064 | 0.016 | 0.083 | 0.09 | -0.022 | -0.047 |

| | SAVAGE | AGGRESSIVE | BEAST | ANIMAL | BRUTE | CREATURE | INHUMAN | INHUMANITY |
|---|---|---|---|---|---|---|---|---|
| BLACK | -0.024 | -0.009 | -0.001 | -0.03 | 0.026 | -0.002 | -0.141 | -0.085 |
| WHITE | -0.036 | -0.054 | 0.016 | -0.027 | -0.02 | 0.006 | -0.147 | -0.093 |
| BLACK_MAN | 0.194 | 0.109 | 0.324 | 0.024 | 0.234 | 0.276 | 0.218 | 0.111 |
| WHITE_MAN | 0.004 | -0.08 | 0.031 | -0.07 | -0.003 | -0.022 | -0.058 | -0.005 |
| MAN | 0.016 | -0.034 | 0.16 | 0.133 | 0.139 | 0.272 | 0.005 | 0.062 |
| WOMAN | -0.015 | -0.026 | 0.094 | 0.096 | 0.082 | 0.215 | -0.005 | 0.016 |
| WHITE_WOMAN | -0.051 | -0.093 | 0.041 | -0.022 | 0.014 | 0.129 | -0.063 | -0.033 |
| BLACK_WOMAN | -0.045 | -0.021 | 0.054 | 0.154 | -0.082 | 0.132 | -0.06 | -0.056 |
| NEGRO | -0.048 | -0.069 | 0.012 | -0.02 | 0.001 | 0.072 | -0.098 | -0.101 |
| CAUCASIAN | -0.048 | -0.017 | -0.007 | -0.031 | -0.011 | 0.077 | -0.054 | -0.089 |
| AFRICAN_AMERICAN | -0.107 | -0.073 | -0.057 | -0.053 | -0.05 | 0.012 | -0.119 | -0.081 |

ACKNOWLEDGMENT

The authors thank anonymous reviewers of previous versions of this manuscript for useful comments and suggestions.

**Christopher L. Dancy** (Member, IEEE) holds a B.S. in computer science and a Ph.D. in information science and technology (with a focus in AI and cognitive science) from the Pennsylvania State University, University Park.

He is an associate professor in the department of Computer Science at Bucknell University. He also holds is an affiliate faculty in the Critical Black Studies department at Bucknell University. His research is at the intersection of AI, computational cognitive science, and Black Studies. He studies how physiological, affective, cognitive, and social processes interact. He uses computational physio-cognitive models to study these processes, as well as what these interactions mean for creating AI systems, especially with respect to critical inspection of the connection of Blackness to the design, development, and use of those systems

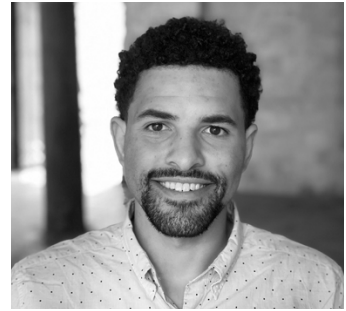

Dr. Dancy has been Chair of the Behavioral Representation in Modeling and Simulation (BRiMS) society, a Program Chair for the SBP-BRiMS conference, and is currently on the advisory board for the Griot Institute for the Study of Black Lives & Cultures at Bucknell University.

**P. Khalil Saucier** is Chair and Professor of Critical Black Studies at Bucknell University (USA) and his thinking, writing, and teaching explore the construction of blackness as the central danger around which western society composes itself. He is author, co-author, editor and co-editor of several books. Those most germane to the study of blackness and artificial intelligence include two edited volumes with Tryon Woods, Conceptual Aphasia in Black: Displacing Racial Formation Theory (Lanham, MD: Lexington Books, 2016) and On Marronage: Ethical Confrontations with Anti-Blackness, (Trenton, NJ: African World Press, 2015). His essays can be found in the Journal of Black Studies, Critical Sociology, Black Studies Papers and more.